\documentclass[aps,preprint,amssymb,12pt,floatfix]{revtex4}
\usepackage{subfigure}
\usepackage{graphicx}
\usepackage{rotating} 
\usepackage{color}
\usepackage{rotating}
 \usepackage{amsmath,amsthm}
\usepackage{epstopdf}
\begin{document}
\title{Signaling Networks and Dynamics of Allosteric Transitions in Bacterial Chaperonin GroEL: Implications for Iterative Annealing of Misfolded Proteins}
\author{D. Thirumalai}
\affiliation{Department of Chemistry, University of Texas at Austin, Texas 78712, USA}
\author{Changbong Hyeon}
\affiliation{Korea Institute for Advanced Study, Seoul 02455, Korea}
\date{\today}

\begin{abstract}
Signal transmission at the molecular level in many biological complexes occurs through allosteric transitions. 
They describe the responses of a complex to binding of ligands at sites that are spatially well separated from the binding region. We describe the Structural Perturbation Method (SPM), based on phonon propagation in solids,  that can be used to determine the signal transmitting allostery wiring diagram (AWD) in large but finite-sized biological complexes. Applications to the bacterial chaperonin GroEL-GroES complex show that the AWD determined from structures also drive the allosteric transitions dynamically. Both from a structural and dynamical perspective these transitions are largely determined by formation and rupture of salt-bridges. The molecular description of allostery in GroEL provides insights into its function, which is quantitatively described by the Iterative Annealing Mechanism. Remarkably, in this complex molecular machine, a deep connection is established between the structures, reaction cycle during which GroEL undergoes a sequence of allosteric transitions, and function in a self-consistent manner.  
\end{abstract}
\maketitle

\section{Introduction}
Allosteric transitions, which refer to responses at distances several nanometers away from the  binding site of ligands to  multi-domain proteins,  are pervasive in biology and are used as a signaling mechanism at the nanoscale level \cite{Hynes02Cell,Liu16PlosCompBiol,Motlag14Nature}. The classic example is the binding of oxygen to hemoglobin (Hb), which triggers quaternary conformational changes in the latter, as explained in the now classic theory \cite{Monod65JMB,Changeux2012ARB} due to Monod, Wyman, and Changeux (MWC). Since the publication of the MWC theory, there has been growing interest in elucidating the molecular and structural basis of allostery in a large number of signaling molecules, including Hb \cite{Eaton99NSB,Viappiani14PNAS}. Achieving this goal has become possible thanks to advances in experimental methods (X-ray crystallography, Small Angle X-ray scattering,  NMR \cite{Lukin03PNAS,Lisi16ChemRev}, and mass spectroscopy \cite{Dyachenko13PNAS}, and more recently cryo EM \cite{Roh17PNAS}) as well as introduction of a variety of computational models. Several review articles have appeared recently  \cite{Changeux13NatRevMolCellBiol,Marzen13JMB,Gruber16ChemRev}  showcasing the spectacular impact of the concept of allostery in biology.  Although there are arguments that signal transmission could occur without significant conformational changes in the complex \cite{Cooper84EBJ}, in most cases allosteric transitions are accompanied by large structural changes. This is indeed the case  in the example we use to illustrate the general concepts of network of residues involved in signal transmission and the accompanying dynamics of allosteric transitions between two distinct states. 

Our focus here is to describe a few concepts associated with transmission of allosteric signals in the bacterial chaperonin GroEL both from a structural and dynamical perspective. Chaperones, which should be viewed as molecular machines like kinesin or myosins,  have evolved to rescue substrate proteins (SPs) that are otherwise destined for aggregation. The GroEL-GroES chaperonin system \cite{ThirumalaiARBBS01}, which functions out of equilibrium by consuming ATP lavishly \cite{Chakrabarti17PNAS,Goloubinoff17biorxiv}, is a promiscuous nano-machine whose spectacular allosteric transitions during its catalytic cycle allows sufficient number of SPs to reach the folded state in biologically relevant timescale.  Thus, understanding the operation of the chaperonin system in molecular terms is of utmost importance in describing its function both {\it in vitro} and {\it in vivo}. 

GroEL is a homo oligomer with two heptamers that are stacked back-to-back. 
The subunits, which are identical, thus confer GroEL an unusual seven fold symmetry in the resting ($T$ or taut) state. Large scale conformational changes between the allosteric states of GroEL, $T\rightarrow R$ and $R \rightarrow R^{\prime\prime}$ transitions (see Fig. 1 for a schematic of the reaction cycle in a single ring), are triggered solely by ATP binding and hydrolysis. The ATP binding sites are localized in the  equatorial (E) domain in which much of the mass resides. The nature of the reversible $T \leftrightarrow R$ transition was first elucidated in pioneering studies by Yifrach and Horovitz \cite{Horovitz01JSB,Yifrach95Biochem} who also established an inverse relation, predicted using computations \cite{Betancourt99JMB},  between the extent of cooperativity in this transition and the folding rates of SPs \cite{Yifrach00PNAS}. The irreversible R$\rightarrow$R" transition is driven by ATP hydrolysis. In both these transitions strain due to ATP binding and hydrolysis at the catalytic site propagates through a network of inter-residue contacts \cite{Tehver09JMB}, thus inducing  large scale conformational changes. That such changes must occur during the reaction cycle of GroEL is already evident by comparing  the static crystal structures in different allosteric states, such as the T and R" states \cite{SiglerNature97}. However, the static structures do not provide any information about the network of residues that carry allosteric signals, the dynamics of transition between the key states in the GroEL reaction cycle, and most importantly a link to the function of GroEL.

In this perspective, we describe a general computational method, the Structural Perturbation Method (SPM) \cite{Zheng05Structure} to determine a network of residues, referred to as the Allosteric Wiring Diagram (AMD), which is largely responsible for transmitting signals between different regions of the protein. The efficacy of the method is illustrated here using GroEL.  Applications to other systems such as DNA polymerases and myosin motors can be found in \cite{Zheng06PNAS,Zheng09BJ}. We then show, using a technique for studying the dynamics of allosteric transition between two states \cite{Hyeon06PNAS}, that the AWD residues are also involved in the transition between distinct states in GroEL. Finally, we show that the large scale molecular rearrangements that occur during the reaction cycle  are linked, through the Iterative Annealing Mechanism (IAM), to the function of GroEL, which is to assist the folding of SPs. The established connection between AWD and its role in the dynamics of allosteric transitions and function shows, in a profound way, how the GroEL architecture and non-equilibrium effects that occur during the catalytic cycle are linked to function.

\section{Determination of the Allostery Wiring Diagram}
We begin by describing the theoretical basis for the Structural Perturbation Method (SPM) \cite{Zheng05Structure,Zheng06PNAS}, which hinges on two ideas that are well-known in condensed matter physics dealing with propagation of excitations in ordered solids. In general, transmission of signals across the nanoscale structures, such as GroEL-GroES system, must satisfy two requirements. 
(1) At least a portion of the complex must be stiff. More precisely, the network of residues that transmits allosteric signals must be capable of bearing ligand-induced  strain over almost the length of the complex. The need for this requirement can be explained using an analogy to the transmission of local disturbance in regular solids. In this case, vibrations of atoms in lattice sites are carried along the entire sample by phonons. The propagation of excitation is possible  because of the stiffness or rigidity of the solid with long-range order, and cannot occur in liquids. (2) The presence of stiff regions, linked by the network of residues (AWD) in a biological complex, implies that the regions associated with allosteric signaling have lower symmetry than the disordered regions, permitting them  to transmit signals across the complex. Let us explain what is meant by lower symmetry. Consider a protein that is unfolded. This state might be viewed as high symmetry state, like a liquid. But in such a state allosteric signals cannot be transmitted because of the absence persistent order. Folded states are aperiodic with lower symmetry than unfolded states, and hence could act as allosteric states in which binding of a ligand in some part could trigger responses elsewhere. In addition, many of the allosteric proteins contain several domains arranged in a symmetric manner and hence capable of propagating stress. In solids translational symmetry is broken for example, thus lowering the symmetry with respect to liquids with short range order. As a consequence  the ordered state is described by elastic constants. In the same vein finite-sized biological complexes the AWD must accommodate excitations across the length scale of the structure, implying that at least a portion of the complex must be structured, which defines the ``allosteric state." This implies that the allosteric states must have lower symmetry compared to disordered states bearing higher symmetry, which comports well with the general description that the functional states of biological molecules are aperiodic \cite{schrodinger1943life}.

\begin{figure}[ht]
\includegraphics[width=4.0in]{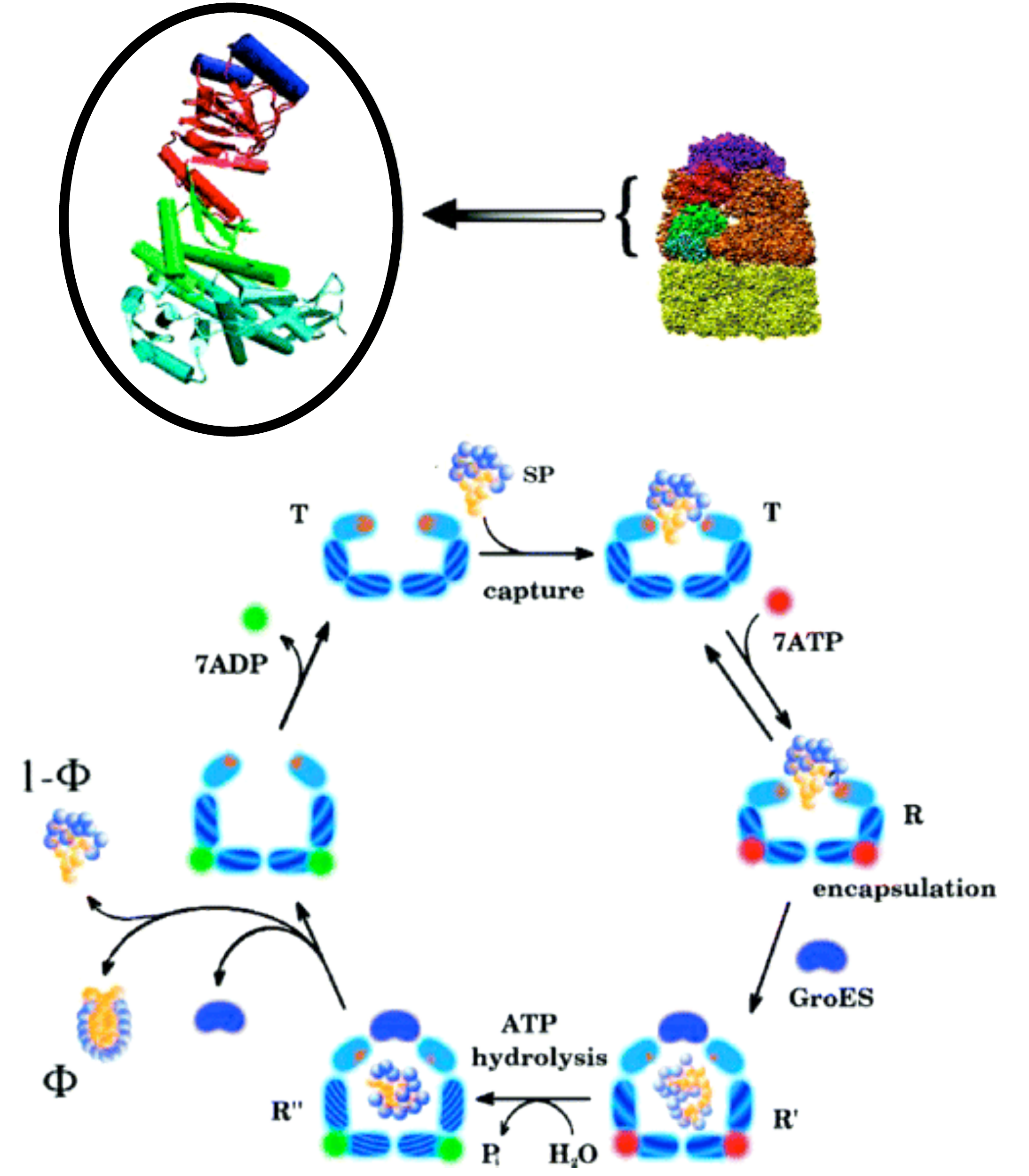}
\caption{On the top right corner is the representation of GroEL-GroES in the post ATP hydrolysis $R^{\prime\prime}$ state \cite{SiglerNature97}. The cartoon representation in the oval corresponds to a single subunit of GroEL. The aqua blue, green and red and blue correspond to E, I, and A domains, respectively. The dark blue cylinders, associated with the A domain, are the H and I helices that recognize the substrate proteins when presented in the misfolded form. The bottom panel is a representation of the coupling of the catalytic cycle and the fate of the SP in the {\it cis} ring. The various steps are: (i) Recognition of the SP when GroEL is in the $T$ state by the helices indicated in dark blue on the top right corner. This is followed by ATP-binding, resulting in the $T \rightarrow R$ transition. GroES binding encapsulates the SP in the central cavity for a very short time during which it can fold with probability $\Phi$ (the native state yield in one hemi-cycle). Following ATP hydrolysis the transition from $R \rightarrow R^{\prime\prime}$ transition. Subsequently, GroES detaches, the inorganic phosphate and ADP are released. In addition, the SP is ejected regardless of whether it has reached the folded state or not, and the cycle begins anew. It should not go unstated that the functional unit is the symmetric complex in which two SP molecules can be processed one in each ring (see \cite{Takei12JBC,Ye13PNAS} for details). 
 \label{RxnCycle}}
\end{figure}

In the second point above we are referring to structural symmetry. The complex that carries allosteric signals must be structured, at least in parts. For GroEL, the entire complex has seven fold symmetry (see Fig. \ref{RxnCycle}). In addition, the subunits also have symmetric arrangements of the individual secondary structural elements, resulting in an aperiodic structure. We should add that the symmetry need not be fully preserved during the allosteric transition. However, all the states involved in signal transmission must be (at least partially) be rigid. 

We use the analogy to phonon propagation in solids to describe the SPM method, first introduced by Zheng and coworkers \cite{Zheng05Structure,Zheng06PNAS} (see for related ideas in biophysics \cite{Kumar15BJ,McLeish15BJ,Townsend15JBC,Flechsig17BJ} and other areas \cite{Yan17PNAS,Rocks17PNAS}) for determining the network of residues that propagate signals upon ligand (ATP and/or SP) binding to specific regions in the complex of interest. In order to determine the AWD, we represent the structure of a given allosteric state (for example the $T$ state in GroEL) as a elastic network of connected springs. 
Usually in the Elastic Network Model (ENM) the structure is represented as a contact map using the $\alpha$ carbon of each residue. A contact implies that the Euclidean distance between the  $\alpha$ carbon atoms of two residues is less than a specific distance. We used a generalization of the standard ENM \cite{Bahar2007COSB,Bahar10ARB} by representing each residue by two beads \cite{Tehver09JMB}, one representing the $\alpha$ carbon and the other the center of mass  of the side chain (SC)  with Glycine being an exception. The center of mass is determined using the side-chain heavy atoms. For Gly, only the $\alpha$ carbon atom is used. 

Following the insightful studies by Bahar and coworkers, who pioneered the applications of ENM to a variety of systems \cite{Bahar05COSB,BaharPRL97,BaharBJ01}, we impose a harmonic potential between all the interaction sites ($\alpha$ carbon atoms and the SCs) that are within a cutoff radius $R_c(\approx 10{\text{ \AA}})$ in the given allosteric structure. In the structure-based elastic network representation of the protein, the potential energy is,
\begin{equation}
E = \frac{1}{2} \sum_{i,j:d_{ij}^0 < R_c} \kappa_{ij} (d_{ij} - d_{ij}^0)^2
\label{ENM}
\end{equation}
where $d_{ij}$  is the distance between the interaction centers $i$ and $j$, $d_{ij}^0$ is the corresponding distance in the native structure, and $\kappa_{ij}$ is the spring constant. The sum is over all the pairs of sites that are in contact in the native conformation. The sites $i$ and $j$ are assumed to be in contact if $d_{ij} < R_c$. The value of $R_c$ is chosen  to ensure  that the B-factors calculated using Eq. \ref{ENM} and the measured values are as close as possible \cite{Zheng07BJ,Tehver08JMB}.  The residue-dependent spring constants, $\kappa_{ij}$, are chosen  to reflect its physical properties.  In the GrOEL applications, we chose $\kappa_{ij}  = \epsilon_{ij}/(\sigma_i/ 2 + \sigma_j/2)^2$ where $\epsilon_{ij}$ is the Betancourt-Thirumalai statistical potential \cite{Betancourt99ProtSci}, which is defined based on contact frequencies as in Miyazawa-Zernigan's \cite{miyazawa1996JMB} and $\sigma_i$ is the van der Waals diameter of the $i^{th}$ residue. 
\\

{\bf Normal Mode Spectrum:}
The first step in the SPM is to first perform a normal mode analysis using the energy
function in Eq. \ref{ENM} in order to generate the spectrum of frequencies for
the normal modes along with the corresponding eigenvectors. Applications of  ENM to a large number of systems including GroEL \cite{Bahar2007COSB,Zheng06PNAS,Zheng07BJ,Zheng09CurrProtScience} have shown that typically  only a few of the lowest-frequency normal
modes are required to characterize the allosteric transitions.  
In order to identify the modes that best describe the transition between two allosteric states, 
$\alpha$  and $\beta$, we
compute the overlap between the vector quantifying the change of states from $\alpha$ to $\beta$ and the $M$-th normal mode (${\bf a}_M$, calculated based on the state $\alpha$), $I_M^{\alpha \rightarrow \beta}$ \cite{Zheng05Structure}. 
The function $I_M^{\alpha \rightarrow \beta}$, which quantifies the extent of overlap between the two vectors, is given by
\begin{equation}
I_M^{\alpha \rightarrow \beta} = \frac{\sum_{i=1}^N \mathbf a_{iM} \Delta  \mathbf r_i}{\sqrt {\sum_{i=1}^N \mathbf a_{iM}^2 \sum_{i=1}^N \Delta \mathbf r_i^2}}
\label{overlap}
\end{equation}
where $N$ is the number of residues in the protein, and $\Delta {\bf r}_i$ is the change in the position of the $i^{th}$ site
between the states $\alpha$ and $\beta$. It follows from Eq. \ref{overlap} that 0$\le I_M \le$1. 
By evaluating Eq.\ref{overlap} for a given $M$, we identify the best overlapping mode that maximizes $I_M^{\alpha \rightarrow \beta}$.
\\

{\bf SPM in practice:}
The extent to which a residue at a given site in a structure responds to a
perturbation far away can be used to assess allosteric coupling. The SPM allows us to quantify the strength of such a coupling  to a mutation at a particular site. The greater the
response (higher $\delta \omega_{iM}$, defined below), the more significant a
specific residue is to a given mode. 

In practice, the SPM probes the response of a normal mode $M$ to a mutation of a residue $i$. In the ENM, perturbation of the spring constant around a site mimics the effect of a mutation. The response to such a perturbation is calculated using,
\begin{equation}
\delta \omega_{iM}^{\alpha \rightarrow \beta} = \frac{1}{2} \sum_{i,j:d_{ij}^0 < R_c} \delta \kappa_{ij} (d_{ij,M} - d_{ij}^0)^2
\end{equation}
\noindent where $\delta \kappa_{ij}$ is the perturbed spring constant and $d_{ij,M} - d_{ij}^0$ is the displacement in residues $i$ and $j$ in the $M^{th}$ mode. Residues  with high $\delta \omega_{iM}$ (large stored elastic energy) constitute the AWD or a network of residues that transmit allosteric signals. We have shown that the AWD residues are also strongly conserved \cite{Zheng06PNAS}, thus underscoring their functional importance.
\\

\begin{figure}[ht]
\includegraphics[width=6.00in]{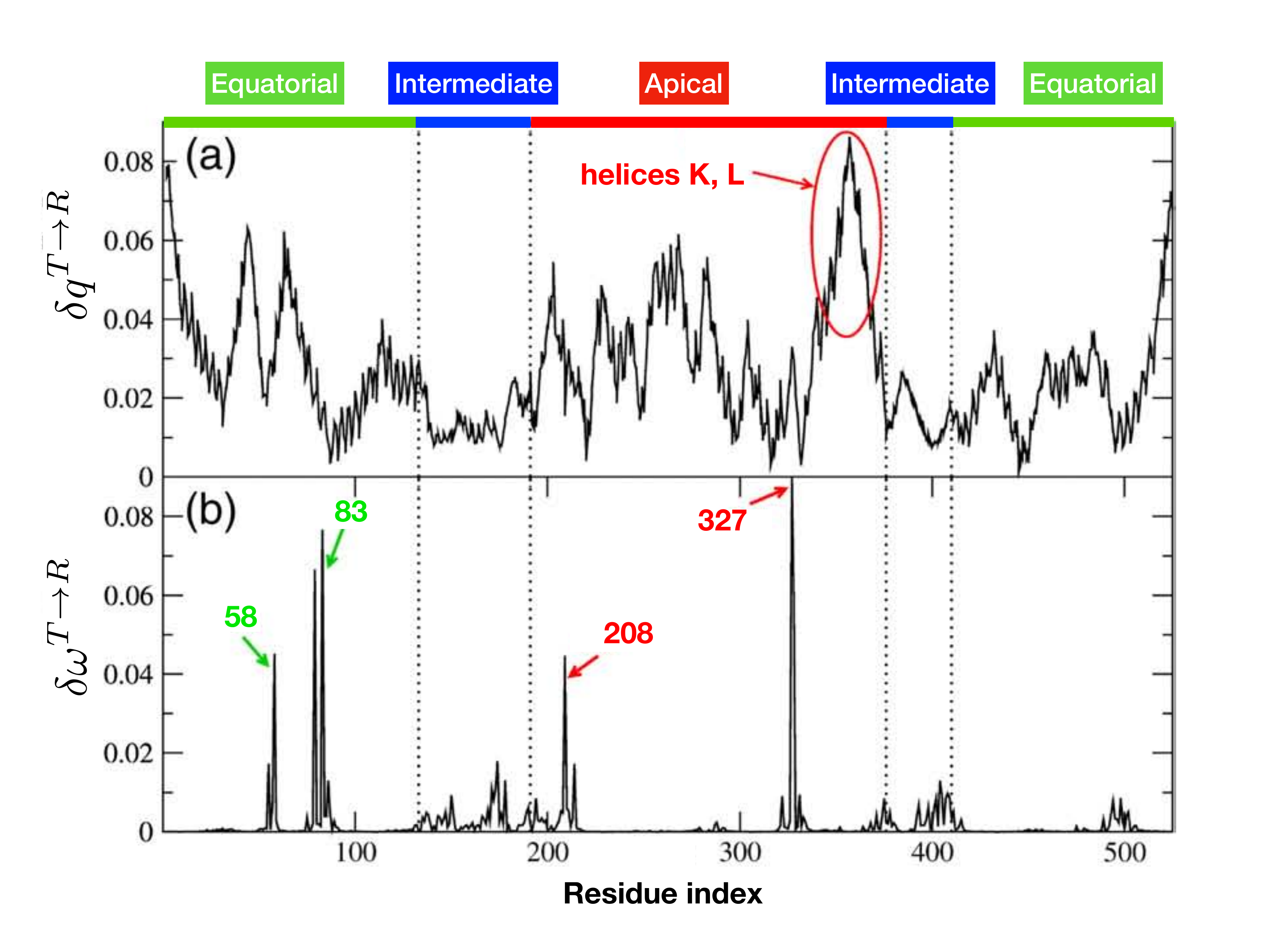}
\caption{SPM results for the GroEL with two adjacent subunits. (a) The amplitudes of motion in the dominant normal modes. The region with the highest amplitude corresponds to helices K and L. (b) Residue-dependent $\delta \omega$ for the dominant modes. The residues with the largest allosteric signal transmitting  values $\delta \omega$ are identified. The figure is adopted from \cite{Tehver09JMB}.
 \label{SPM}}
\end{figure}

\textbf{Asymmetry in the response between subdomains in the $T \rightarrow R$ transition:} The seven subunits of the oligomer of GroEL are identical. However, it is thought that the allosteric responses of each subunit might be asymmetric in the sense that the amplitude of fluctuations in two E or I domains may be different. As a way of illustrating the application of the SPM and to illustrate the asymmetric response we performed normal mode analysis and estimated the effect of perturbation at a specific site on the whole structure using the procedure outlined above. To identify the most significant residues including those at the interface, we constructed two subunits of GroEL in the $T$ state (Fig. \ref{RxnCycle}). The $T \rightarrow R$ transition of a GroEL model, with two adjacent subunits, is best described by two  modes with significant values of the  overlap 0.35) (see Fig. 4a in \cite{Tehver09JMB}). Here a few important points are worth making. (i) The amplitudes of vibration for the two modes, shown in Fig. \ref{SPM}a indicate that there is a noticeable reduction in the fluctuations of the intermediate (I) domain residues. (2) In addition, helices K and L (residues 339-371) show the largest amplitude among the apical (A) domain residues. The SPM result for the modes, displayed in Fig. \ref{SPM},  show that residues D83 and K327 have the largest $\delta \omega$ values.  (3) Notice that this figure also shows the asymmetry in the high $\delta \omega$ values between identical subunits. For example, the high values in both the amplitude of fluctuations and the $\delta \omega$ values in one of the E domain (left side in Fig.\ref{SPM}) is absent in the other. (4) The largest fluctuations in both modes (7 and 13) are localized in the A domain, which as shown below, is also reflected in the dynamics of the allosteric transitions. 

\begin{figure}[ht]
\includegraphics[width=7.00in]{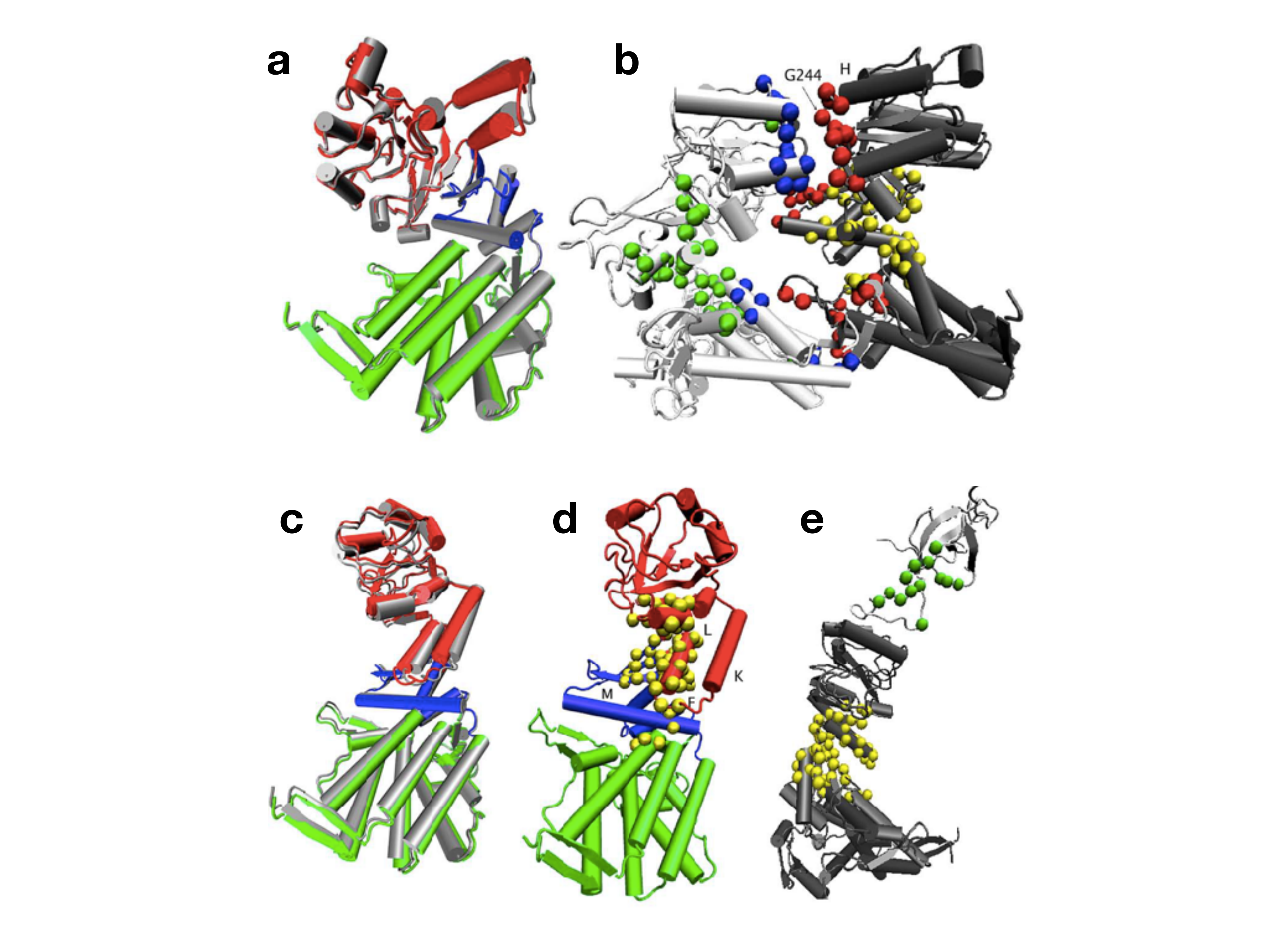}
\caption{llustrating the $T \rightarrow R$ transition and the associated allostery wiring diagram determined using the SPM. (a) Single-subunit structure in the $T$ state. The E, I, and A domains are shown in green, blue, and red, respectively. The motions of the structural elements due to the dominant mode are in gray. (b) Structure of two adjacent subunits of GroEL (the chains are shown in dark and light gray) in the $T$ state. The  residues in the AWD are highlighted in color. The critical interface residues are in red and blue, and the other hot-spot residues are in yellow and green. (c) Same as (a) except this describes the $R \rightarrow R^{\prime\prime}$ transition. (d) The AWD  for the transition from the $R^{\prime\prime}  \rightarrow T$ state are in yellow. Helices K, L, F, and M  are labeled. The domains are colored as in (c). (e) GroEL (dark gray)ÐGroES (light gray) model. The AWD  is shown in yellow (GroEL) and green (GroES). This figure provides insights into the residues that signal the disassociation of GroES, a key event in the function of the chaperonin.
\label{AWD}}
\end{figure}

By mapping the hot-spot residues (listed in \cite{Tehver09JMB}) onto their structures, we find that 33 of the 85 hot-spot residues of chain H (per the chain labeling in the PDB structure 1AON) and 24 of the 62 hot-spot residues of chain I belong to the inter-subunit interface. We define interface residues as residues that make at least one contact with a residue in the adjacent subunit. The interface hot-spot residues, highlighted in blue and red in Fig. \ref{AWD}b, show that the large number of interface residues in the AWD is the possible foundation for the strong intra ring positive cooperativity.


\section{Dynamics of Allosteric Transitions}

The findings based on the SPM highlight the most probable AWD driving the transition between two allosteric states. To better understand the allosteric transitions of GroEL particle at the microscopic level, we performed multiple sets of Brownian dynamics simulations using the self-organized polymer (SOP) model \cite{HyeonBJ07,Hyeon06Structure}. The SOP model uses  a united atom representation lumping the heavy atoms in each amino acid  into a single interaction center. 
This novel coarse-grained model has been used to make several important contributions to the theoretical biological physics in the area of RNA \cite{HyeonBJ07,Hyeon06Structure,Lin08JACS} and protein dynamics \cite{Hyeon06PNAS,Mickler07PNAS,Hyeon07PNAS,Hyeon07PNAS2,Hyeon09PNAS}. 
Here, we describe the simulations \cite{Hyeon06PNAS} used to identify key events in the transition between $T$ and $R$ as well as $R$ and $R^{\prime\prime}$ transitions.
\\

{\bf Overview of the Dynamics of $T\rightarrow R$ and $R \rightarrow R^{\prime\prime}$ transitions:} In the $T\rightarrow R$ transition triggered by ATP binding, the A domains undergo counterclockwise motion mediated by a multiple salt-bridge switch mechanism at the interfaces of the seven subunits, where the salt-bridges are defined for the non-covalent contact made between oppositely charged residues, e.g., between Arg (R), Lys (K) and Asp (D), Glu (E).  The $T \rightarrow R$ transition is accompanied by a series of breakage and formation of salt-bridges. 
The initial event in the $R \rightarrow R^{\prime \prime}$ transition, during which GroEL rotates clockwise, involves a dramatic outside-in concerted movement of helices K and L. 
The outside-in movement of helices K and L, exerting a substantial strain on the GroEL structure, induces the 90$^o$  clockwise rotation and 40$^o$ upward movement of A domain. 
Such a large scale rotation of helices K and L is an entirely new finding using SOP model simulations \cite{Hyeon06PNAS}, which provided a basis for understanding the origin of the change in polarity of the GroEL cavity. In both the transitions, considerable heterogeneity is found in the transition pathways, as discussed below. 
In what follows, we provide further details of the allosteric dynamics of GroEL gleaned from simulations using the SOP model. 
\\

\begin{figure}[ht]
\includegraphics[width=6.00in]{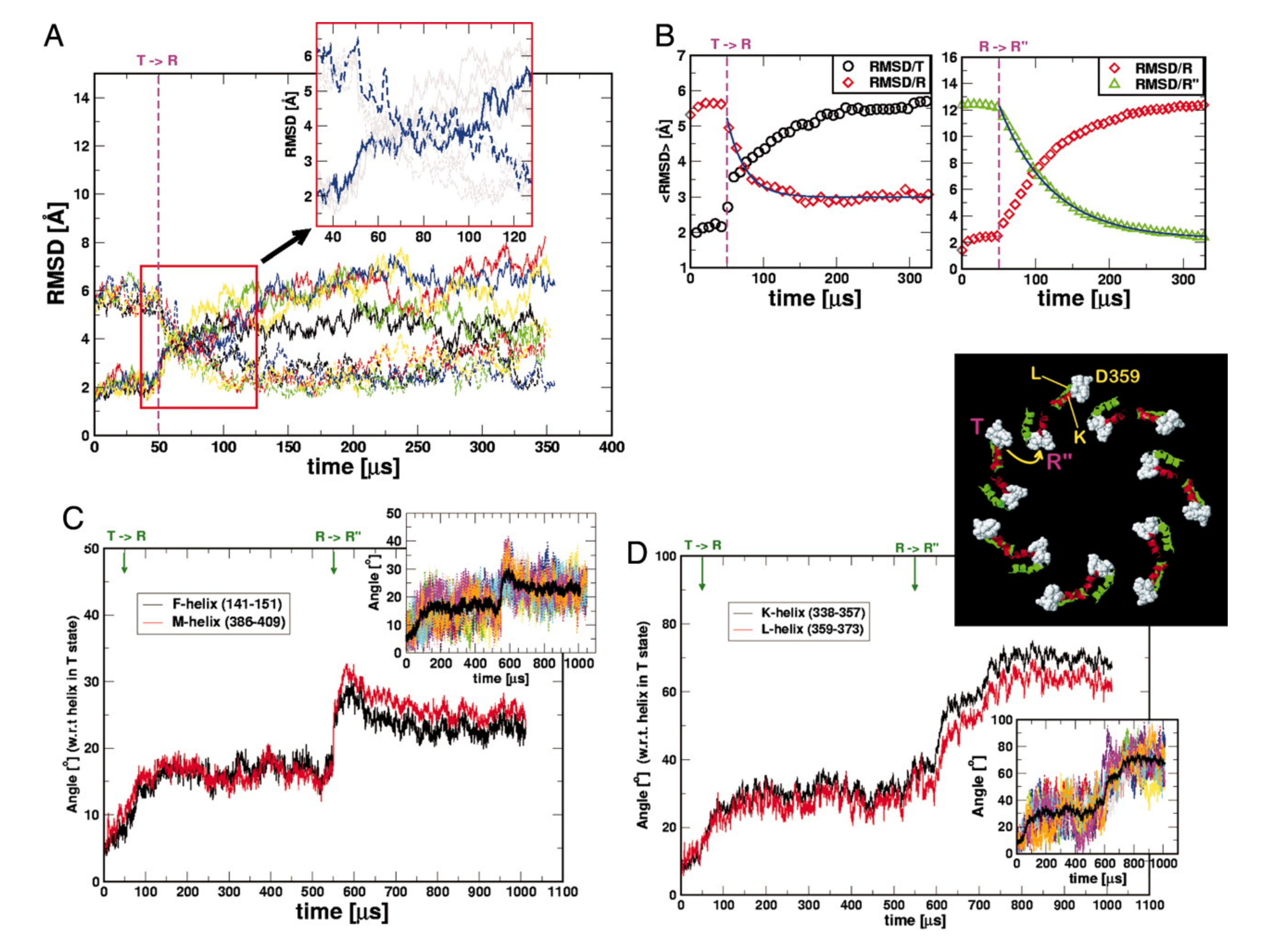}
\caption{
Root-Mean-Squared-Distance (RMSD) as a function of time. 
{\bf A.} Time-dependence of the RMSD  of a few individual molecules are shown for the $T\rightarrow R$ transition. 
Solid (dashed) lines are for RMSD/$T$ (RMSD/$R$)  (RMSD calculated with respect to the $T$ ($R$) state). 
The enlarged inset gives an example of a trajectory, in blue, that exhibits multiple passages across the transition region. 
{\bf B.} Ensemble averages of the RMSD for the $T\rightarrow R$ (top) and $R\rightarrow R^{\prime\prime}$ (bottom) transitions. 
The solid lines are exponential fits to RMSD/$R$ and RMSD/$R^{\prime\prime}$  relaxation kinetics. {\bf C.} Time-dependent changes in the angles (measured with respect to the $T$ state) that  F, M helices make during the $T\rightarrow R\rightarrow R^{\prime\prime}$ transitions. The inset shows the dispersion of individual trajectories for the F-helix  with the black  line being the average.
{\bf D.} Time-dependent changes in the angles (measured with respect to the $T$ state) that  K, L helices make during the $T\rightarrow R\rightarrow R^{\prime\prime}$ transitions. The inset on the top shows the structural changes in  K, L helices during the $T\rightarrow R\rightarrow R^{\prime\prime}$ transitions. For clarity, residues 357-360 are displayed in space-filling representation in white.  The dispersion of individual trajectories for the K-helix is shown in the inset at the bottom.  The black  line is the average. In (C) and (D) $\theta = \cos^{-1}(u(0)\cdot u(t))$.
\label{RMSD_and_angle.fig}}
\end{figure}

{\bf Global $T \rightarrow R$ and $R \rightarrow R^{\prime \prime}$ transitions follow two-state kinetics:}  
The overall conformational change that occurs during the allosteric transitions  can be quantified using a global coordinate characterizing the structural overlap with respect to a reference conformation. 
Time-dependent changes in the root-mean-squared-distance (RMSD) with respect to a reference state ($T$, $R$, or $R^{\prime\prime}$), from which a specific allosteric transition commences  (Fig.\ref{RMSD_and_angle.fig}), 
differ from one trajectory to another, reflectsing the heterogeneity of the underlying dynamics (Fig.\ref{RMSD_and_angle.fig}-A). 
Examination  of the RMSD for a particular trajectory in the transition region (Fig.\ref{RMSD_and_angle.fig}-A inset) shows that GroEL particle recrosses the transition state (TS).  Assuming that RMSD is a reasonable representation of the structural changes during the allosteric transitions, we find that GroEL spends a substantial fraction of time (measured with respect to the first passage time for reaching the R state starting from the T state) in the TS region during the $T \rightarrow R$ transition.  After an initial increase (decrease) with respect to the $T$ ($R$) state the RMSD changes non-monotonically in the transition region, 
which suggests that the  transition state ensemble (TSE) connecting the two end states is broad (details follow).  
By averaging over fifty individual trajectories, we find that the ensemble average of the time-dependence of RMSD for both the  $T\rightarrow R$ and 
$R\rightarrow R''$ transitions follow single exponential kinetics, which clearly obscures the molecular heterogeneity observed in individual trajectories.  
Despite such complex dynamics at the individual molecule level, the ensemble average allosteric transition kinetics can be approximately described by a two-state model.  
Unlike the global dynamics characterizing the overall  motion of GroEL, 
the local dynamics describing the formation and rupture of key interactions associated with GroEL allostery cannot be described by the two-state kinetics, which clearly not only reflects the heterogeneity but also shows a certain hierarchy in the dynamics of allosteric signaling at the molecular level (see below). \\ 

{\bf $T\rightarrow R$ transition is triggered by downward tilt of helices F and M 
followed by a multiple salt-bridge switching mechanism: }
Several residues in helices  F  and  M in the I domain (Fig. \ref{RxnCycle}) interact  with the nucleotide-binding sites in the E domain thus creating a tight nucleotide binding pocket.  
The tilting of F, M helices by $\sim 15^o$ closes the nucleotide-binding sites, the residues around which are highly conserved \cite{Brocchieri00ProtSci,Stan03BioPhysChem}.
Since the $T \rightarrow R$ transition involves the formation and breakage of intra- and inter-subunit contacts, 
we simulated two adjacent, interacting subunits, which allowed us to dissect the order of events.

\begin{figure}[ht]
\centering
\includegraphics[width=6.0in]{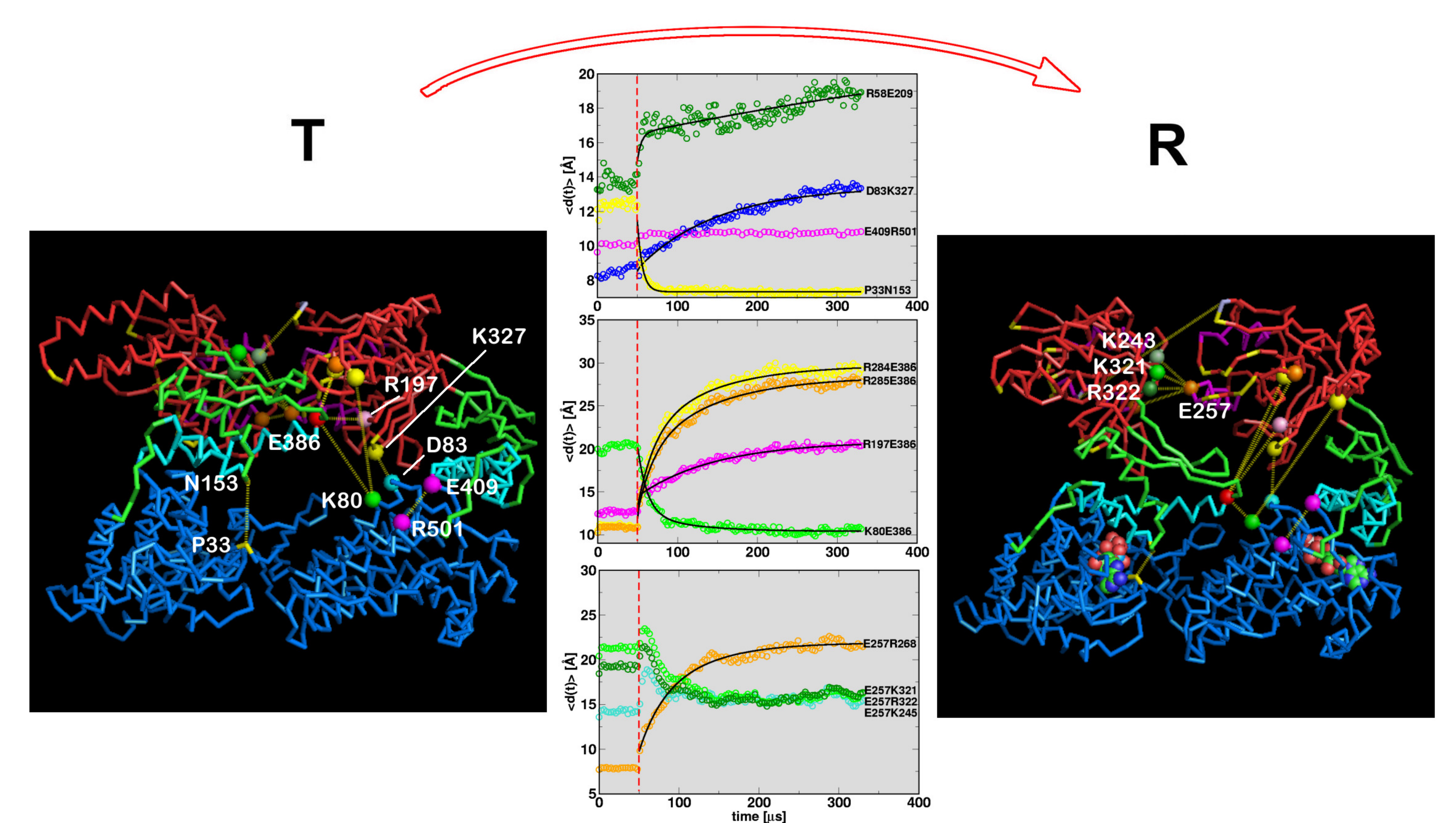}
\caption{$T\rightarrow R$ GroEL dynamics monitored using of two interacting subunits. 
Side views from outside to the center of the GroEL ring and top views are presented for the $T$ (left panel) and $R$ (right panel) states. 
Few residue pairs  are annotated and connected with dotted lines.  
The ensemble average kinetics of a number of salt-bridges and contacts between few other residues are shown in the middle panel.  
Relaxation dynamics of distance between some of two residues of interest are fitted to the multi-exponential kinetics.  The fits for the breakage and formation of the various contacts can be found in \cite{Hyeon06PNAS}. Note that the definition of salt-bridge in the SOP model with a single bead for each residue, located at the $C_{\alpha}$ position, is longer than it would be in an all-atom representation by a few \AA. 
\label{saltbridgeanalysisTR.fig}}
\end{figure}

(i) The ATP-binding-induced concerted downward tilt of the  F, M helices is the earliest event \cite{KarplusJMB00} $T \rightarrow R$ transition. 
The changes in the angles that F and M helices make
with respect to their orientations in the $T$ state occur in concert  (Fig.\ref{RMSD_and_angle.fig}-C). 
At the end of the $R \rightarrow R^{\prime \prime}$ transition the helices have tilted on average by about $25^o$ in all  (Fig.\ref{RMSD_and_angle.fig}-C). 
The downward tilt of the F and M helices narrows the entrance to the ATP binding pocket as evidenced by the rapid decrease in the distance  
between P33  and N153 (Fig.\ref{saltbridgeanalysisTR.fig}).  
The contact number of N153 increases substantially as a result of loss in accessible surface area during the $R \rightarrow R^{\prime \prime}$ 
transition \cite{Stan03BioPhysChem}. 
In the $T$ state E386, at the tip of M helix, forms inter-subunit salt-bridges with R284, R285, and R197, which are disrupted and forms a new intra-subunit salt-bridge with K80 (see the middle panel in Fig.\ref{saltbridgeanalysisTR.fig}). 
The tilting of M helix must precede the formation of inter-subunit salt-bridge between E386 and K80.

(ii) The rupture of the intra-subunit salt-bridge  D83-K327 occurs nearly simultaneously with the
disruption of the E386-R197 inter-subunit interaction with relaxation time on the order of $\tau\approx 100$ $\mu s$ (the blue kinetic curve at the top panel in the middle of Fig.\ref{saltbridgeanalysisTR.fig}). 
K80-E386 salt-bridge is formed around the same time as the rupture of R197-E386 interaction.
 In the $T \rightarrow R$ a network of salt-bridges breaks and new ones form (see below). 
At the residue level, the reversible formation and disruption of D83-K327 salt-bridge, in concert with
the inter-subunit salt-bridge switch associated with E386 \cite{Ranson01Cell} and E257 \cite{Stan05JMB,Danziger06ProtSci}, are among the most significant  events that dominate the $T \rightarrow R$ transition. 

The coordinated global motion is orchestrated by a multiple salt-bridge switching mechanism.  
The movement of the  A domain results in the dispersion  of the SP binding sites and  also  leads to the rupture of the  E257-R268 inter-subunit salt-bridge. The kinetics of breakage of the E257-R268 salt-bridge is distinctly non-exponential (the orange kinetic curve at the bottom panel in the middle of Fig.\ref{saltbridgeanalysisTR.fig}).  
It is very likely that the dislocated SP binding sites maintain their stability through the inter-subunit salt-bridge formation between the A domain residues.  To maintain the stable configuration in the R state, E257 engages in salt-bridge formation with positively charged residues that are initially buried at the interface
of the inter-A domain in the T state.  Three positively  charged residue at the interface of the A domain in the R state, namely, 
K245, K321, and R322 are the potential candidates for the salt-bridge with E257.  During the $T \rightarrow R$ transitions E257 interacts partially with K245, K321, and R322 as evidenced by the decrease in their distances (the last panel in the middle column of Fig.\ref{saltbridgeanalysisTR.fig}). The distance between E409-R501 salt-bridge, holding the  I and E domains, remains intact at a distance  $\sim 10$ \AA\ throughout the whole allosteric transitions. 
This salt-bridge and two others (E408-K498 and E409-K498) might be important for enhancing positive intra-ring cooperativity and for stability of the chaperonins.  Indeed, mutations at sites E409 and R501 alter the stability of the various allosteric states \cite{Aharoni97PNAS}.  
In summary, we find dynamic changes in the network of salt-bridges coordinate the $T\rightarrow R$ transition.  

It is worth emphasizing that the order of events, described above, is not followed in all the trajectories.   
Each GroEL molecule follows somewhat different pathway during the allosteric transitions which is indicated by the considerable dispersion in the dynamics. Recent cryo-EM study \cite{Roh17PNAS} has shown that there is considerable heterogeneity in the conformations of the various allosteric states. It, therefore, stands to reason that the dynamics connecting the allosteric states will be likewise heterogeneous, which would support our findings from over a decade ago \cite{Hyeon06PNAS}. 
\\

{\bf  $R\rightarrow R^{\prime\prime}$ transition involves a spectacular outside-in movement of K and L helices accompanied by inter-domain salt-bridge formation K80-D359: }
The dynamics of the irreversible $R\rightarrow R''$  transition is propelled by substantial movements in the A domain helices K and L  
 that drive the dramatic conformational change in GroEL resulting in doubling of the volume of the cavity. 
The   $R \rightarrow R^{\prime\prime}$ transition also occur in stages.

\begin{figure}[ht]
\centering
\includegraphics[width=6.0in]{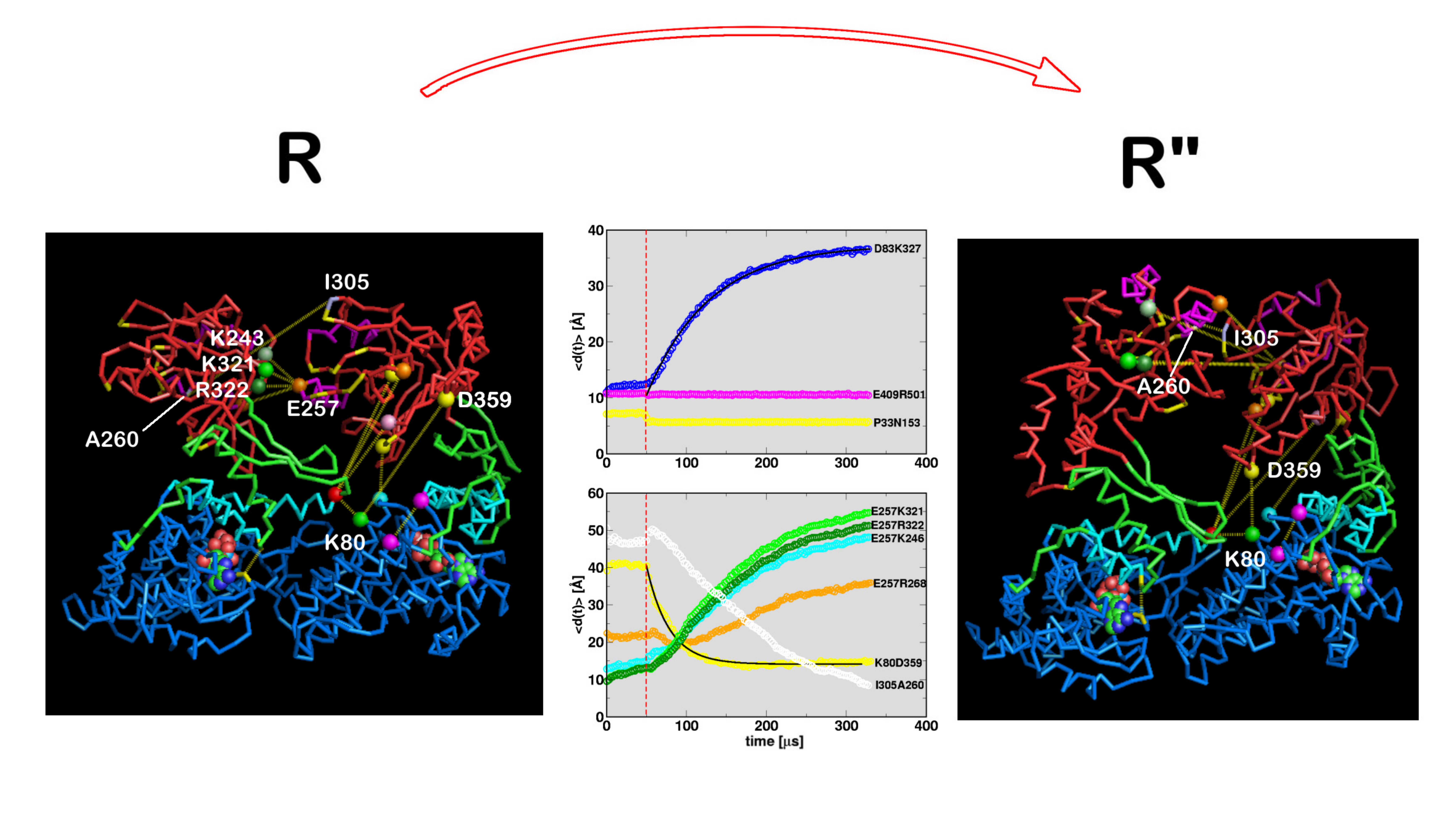}
\caption{
Dynamics of the $R\rightarrow R^{\prime\prime}$ transition using two-subunit SOP model simulations. The dynamics along one trajectory are shown in Fig. S4 in the Supplementary Information.  Intra-subunit salt bridges (or residue pairs) of interest (D83-K327, E409-R501, P33-N153) are plotted on the top panel, and inter-subunit saltbridges  of interest (E257-K246, E257-R268, E257-K321, E257-R322, I305-A260) are plotted on the bottom panel. For emphasis, the dynamics of K80-D359 saltbridge, that provides a driving force to other residue dynamics, is highlighted on the bottom panel.
\label{saltbridgeanalysisRRpp.fig}}
\end{figure}

(i) Upon ATP hydrolysis the F, M helices rapidly tilt by an additional $10^o$ (Fig.\ref{RMSD_and_angle.fig}-C). Nearly simultaneously  there is a small  reduction in P33-N153 distance (7 \AA $\rightarrow$ 5 \AA) (see top panel in Fig.\ref{saltbridgeanalysisRRpp.fig}). These relatively small changes are the initial events in the $R \rightarrow R^{\prime\prime}$ transition. 

(ii) In the subsequent step, the A domain undergoes significant conformational changes that are most vividly captured by the outside-in concerted movement of helices K and L.  These two helices, that tilt by about $30^o$ during the $T \rightarrow R$ transition, further rotate by an additional $40^o$ when the $R \rightarrow R^{\prime\prime}$ transition occurs (Fig.\ref{RMSD_and_angle.fig}-D). In the process, a number of largely polar and charged residues that are exposed to the exterior in the $T$ state line the inside of the cavity in the $R^{\prime\prime}$ state, making the interior of GroEL polar. The outside-in motion of K and L helices leads to an inter-domain salt-bridge K80-D359 whose $C_{\alpha}$ distance changes rapidly   
from about 40 \AA\ in the $R$ state to about 14 \AA\ in the $R^{\prime \prime}$  (Fig.\ref{saltbridgeanalysisRRpp.fig}). 

The wing of the A domain that protrudes outside the GroEL ring in the $R$ state moves inside the cylinder. 
The outside-in motion facilitates the K80-D359 salt-bridge formation which in turn orients the position of the wing.  
The  orientation of the A domain's wing inside the cylinder exerts a substantial strain (data not shown) on the 
GroEL structure. 
To relieve the strain,  the A domain is forced to undergo  
a dramatic 90$^o$ clockwise rotation and 40$^o$ upward movement with respect to the $R$ state. 
As a result, the SP binding sites (H, I helices, colored blue in Fig. \ref{RxnCycle}) are oriented in 
the upward direction.
Before the strain-induced alterations are possible the 
distance between K80 and D359 decreases drastically from that in R state (middle panel in Fig. \ref{saltbridgeanalysisRRpp.fig}). 
The clockwise motion of the A domain occurs only after the formation of salt-bridge between K80 and D359.   
The formation of the salt-bridge K80-D359 is followed by the disruptions of several salt-bridges between the inter-A domain residues, K245, E257, R268, K321, and R322 (Fig.\ref{saltbridgeanalysisRRpp.fig}). 
Formation of contact between I305 and A260 (a binding site for substrate proteins), an  inter-subunit residue pair located at the interface of two adjacent A domains in the  $R^{\prime\prime}$ state, occurs extremely slowly  compared to others.  
The non-monotonic and lag-phase kinetics observed in the rupture and formation of a number of contacts suggests that intermediate states must exist in the pathways connecting the $R$ and $R^{\prime \prime}$ states . 

 The clockwise rotation of A domain, 
orients the domain in the upward direction so as to permit the binding of the mobile loop of GroES. 
Hydrophobic interactions between SP binding sites and GroES drive the  $R\rightarrow R^{\prime\prime}$ transition. 
The hydrophilic residues, hidden on the side of A domain in the $T$ or the $R$ state, are now exposed to form an interior surface of the GroEL. 
The E409-R501 salt-bridge formed between I and A domains close to the $\gamma-P_i$ binding site 
is maintained  throughout the allosteric transitions including in the transition state  \cite{Aharoni97PNAS}.
 \\

\begin{figure}[ht]
\includegraphics[width=7.00in]{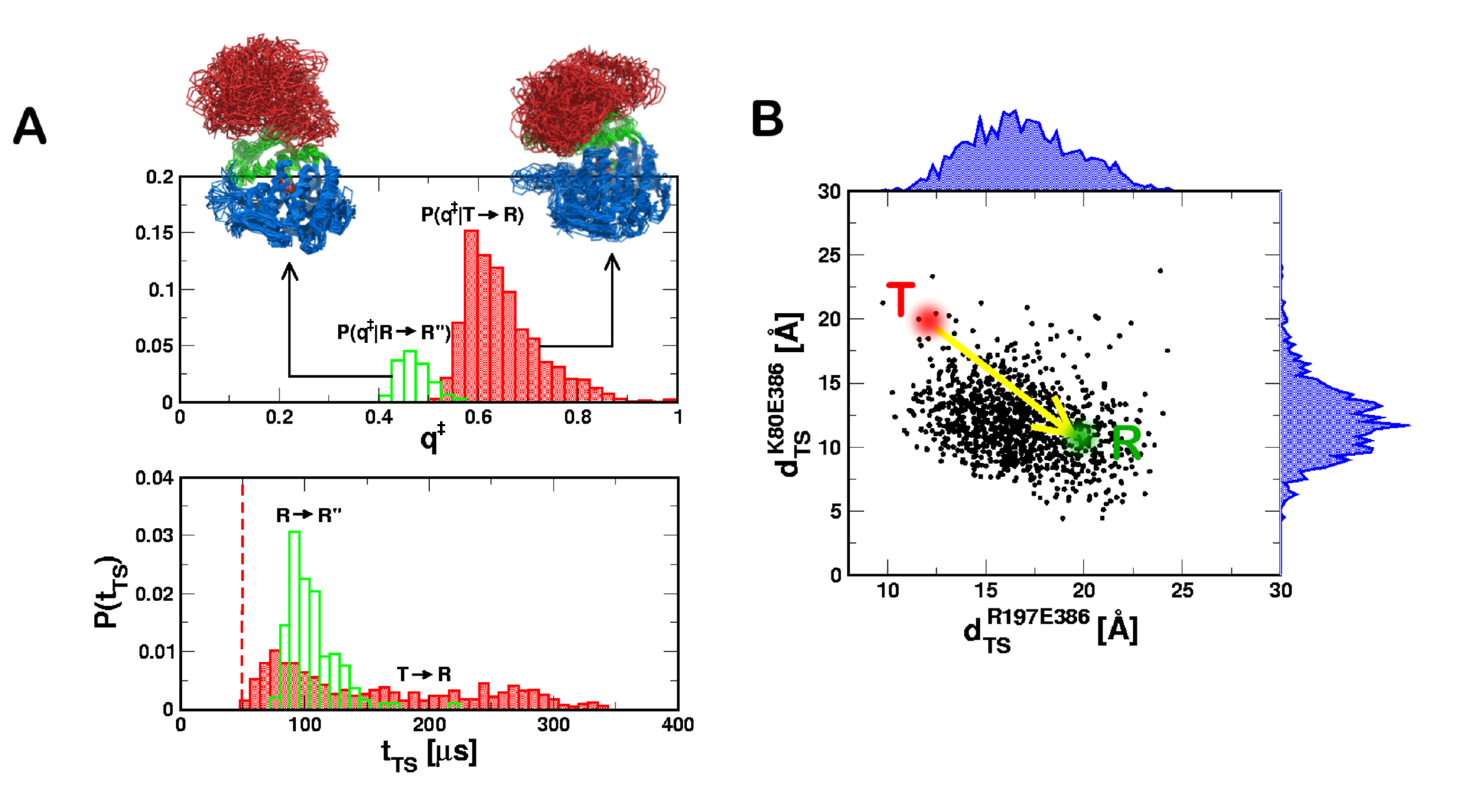} 
\caption{
Transition state ensembles (TSE). 
{\bf A}. TSEs are represented in terms of distributions $P(q^{\ddagger})$ where  
$q^{\ddagger}\equiv\frac{\Delta^{\ddagger}-\min(\text{RMSD}/X)}{\max(\text{RMSD}/X)-\min(\text{RMSD}/X)}$.  
Histogram in red gives $P(q^{\ddagger})$  for $T\rightarrow R$ (red) and the data in green are 
for the $R\rightarrow R^{\prime\prime}$ transitions.   
For $T\rightarrow R$, $X=R$, $\min(\text{RMSD}/X)=1.5$ \AA\ and $\max(\text{RMSD}/X)=8.0$ \AA. 
For $R\rightarrow R''$, $X=R^{\prime\prime}$, $\min(\text{RMSD}/X)=1.5$ \AA\ and $\max(\text{RMSD}/X)=14.0$ \AA. 
To satisfy conservation of the number of molecules  
the distributions are normalized using
$\int dq^{\ddagger}\left[P(q^{\ddagger}|T\rightarrow R)+P(q^{\ddagger}|R\rightarrow R")\right]=1$. 
 Twenty overlapped TSE structures  for the two transitions are displayed.   
In the bottom panel, the distributions of $t_{TS}$ that satisfy  $\delta^{\ddagger} < 0.2$ \AA, 
are plotted for the $T\rightarrow R$ and the $R\rightarrow R''$ transitions. 
{\bf B}. TSE for the $T\rightarrow R$ transition represented by the pair of two salt-bridge distances $(d^{R197-E386}_{TS},d^{K80-E386}_{TS})$ (black dots). 
The equilibrium distances 
$(\langle d^{R197-E386}_{TS}\rangle,\langle d^{K80-E386}_{TS}\rangle)$ in the $T$ and the $R$ states are shown using the red and the green dots, respectively. 
The distance distributions for the TSE are shown in blue.
\label{TSE.fig}}
\end{figure}

{\bf Transition state ensembles (TSEs) connecting the allosteric states are broad: }
The structures of the TSEs 
connecting the $T$, $R$, and $R^{\prime \prime}$ states are obtained using RMSD as a surrogate reaction coordinate.  
We assume that, for a particular trajectory,  the TS location is reached when  $\delta^{\ddagger} = |(\text{RMSD}/T)(t_{TS})-(\text{RMSD}/R)(t_{TS})|<r_c$
where $r_c = 0.2$ \AA, and $t_{TS}$ is the time at which  $\delta^{\ddagger} < r_c$. 
Letting the value of RMSD at the TS 
be $\Delta^{\ddagger} = 1/2\times|(\text{RMSD}/T)(t_{TS})+(\text{RMSD}/R)(t_{TS})|$ the
distributions $P(\Delta^{\ddagger})$ for $T \rightarrow R$ and $R \rightarrow R^{\prime\prime}$ transitions are broad (see Fig. S3 in the Supplementary Information). 
If $\Delta^{\ddagger}$ is normalized by the RMSD between the two end point structures to produce a Tanford $\beta$-like parameter $q^{\ddagger}$ (see caption to Fig. \ref{TSE.fig} for definition), 
we find that the width of the TSE for the $R\rightarrow R^{\prime\prime}$ is less than the $T \rightarrow R$ transition (Fig. \ref{TSE.fig}-A).  
The mean values of $q^{\ddagger}$ for the two transitions show that the most probable TS is located close to the $R$ states in both $T \rightarrow R$ and $R \rightarrow R^{\prime\prime}$ transitions.

Disorder in the TSE structures (Fig. \ref{TSE.fig})  is largely localized in the A domain, which once again shows that the substructures in this domain partially unfold as the barrier crossings occur.  
By comparison  the E domain remains more or less structurally intact even at the transition state, suggesting that the relative immobility of 
this domain is crucial to the function of this biological namomachine \cite{ThirumalaiARBBS01}. It is most likely the case that the E domain is the anchor for force transmission to the SP, thus partially unfolding it, as the reaction cycle proceeds. 
The dispersions in the TSE are also reflected in the heterogeneity of the distances between various salt-bridges in the transition states.  

\section{Functional Implications - Iterative Annealing Mechanism (IAM) \cite{Todd96PNAS}:} 
The dynamics clearly reveals that by breaking a number of salt bridges, over a hierarchy of time scales, the volume of the central cavity expands dramatically, expanding by  two fold. More importantly, in the process the interaction between the SP and GroEL changes drastically. Upon ensnaring the SP the SP-GroEL complex is (marginally) stabilized predominantly by hydrophobic interactions. However, during the subsequent ATP-consuming and irreversible R$\rightarrow$R" transition, not only does the volume double but also the microenvironment of the SP is largely polar. This occurs because of the remarkable nearly 180$^o$ rotations of helices K and L that results in the inner cavity of the GroEL. Thus, during a single catalytic cycle the microenvironment that the SP is subject to changes from being hydrophobic to polar. This change is the annealing mechanism of GroEL that places the SP stochastically from one region, in which the misfolded SP is trapped, to another region from which it could with some probability reach the folded state. The cycle of hydrophobic to polar change takes place with each catalytic cycle, and hence the GroEL-GroES machine iteratively anneals the misfolded SP enabling it to fold.

The physical picture of the IAM, described above qualitatively, whose molecular origin is illustrated by the GroEL allostery has been translated into a a set of kinetic equation with the express purpose of quantitatively describing the kinetics of chaperonin-assisted folding of stringent {\it in vitro} substrates, such as Rubisco \cite{Tehver08JMB}. According to theory of IAM (see Fig. \ref{RxnCycle}) in each cycle, corresponding to the completion of $T\rightarrow R$ and $R\rightarrow R^{\prime\prime}$ transitions, the SP folds by the Kinetic Partitioning Mechanism (KPM) \cite{Guo95BP}. The KPM shows that a fraction, $\Phi$, referred to as the partition factor, reaches the native state. In the context of assisted folding it implies that with each round of folding the fraction of folded molecules is $\Phi$ and the remaining fraction gets trapped in one of the many misfolded structures. After $n$ such cycles or iterations the yield of the native state is,
\begin{equation}
\Psi = 1 - (1 - \Phi)^n
\end{equation}

We illustrate the success of the IAM theory first extracting the key parameters by fitting the kinetic equations to experiments on data. The fits at various GroEL concentration, with a fixed initial concentration of Rubisco, is excellent. Remarkably, for Rubisco the partition factor $\Phi \approx 0.02$, which means that only about 2\% of the SP reaches the folded state in each cycle. The key parameter in the IAM is the rate ($k_{R" \rightarrow T}$) of resetting the machine after ATP hydrolysis, which involves release of inorganic phosphate and ADP, to the taut ($T$) state that can again recognize the SP to start assisted folding. Thus, by maximizing $k_{R" \rightarrow T}$ the native state yield can be optimized in a fixed time. It naturally follows that if the all purpose wild type GroEL has evolved to perform optimally then any mutant of GroEL would produce less of the native fold for a specified time. This consequence of IAM could be tested using data on the ability of GroEL and mutants to rescue the folding of mitochondrial Malate Dehydrogenase (mtMDH) for which experiments have been carried out by Lund and coworkers \cite{Sun2003JMB}. With $k_{R"\rightarrow T}$ as the only parameter the IAM predictions match quantitatively with experiments not only for mtMDH but also citrate synthase \cite{Tehver08JMB}.

\section{Conclusions:}
From the functional perspective it is now firmly established that the IAM is the only theory that explains all the available kinetic data quantitatively \cite{Tehver08JMB}. The theory mandates that with each cycle only a very small fraction (for stringent SP such as Rubisco) reaches the native state. As shown here, during each cycle, GroEL and GroES undergo large scale conformational changes. During each cycle the microenvironment of the inner cavity changes (annealing function), ATP is hydrolyzed, ADP, inorganic phosphate, and the SP (folded or not) are ejected. In the process the machine is reset to the starting taut state for the cycle to begin anew. The release of ADP, which is accelerated by about a hundred fold in the presence of SP \cite{Ye13PNAS}, requires signaling spanning over 100 \AA\, a remarkable example of allosteric communication! We have argued that transmission of signals across such a large distance requires stiffness in at least certain regions of the GroEL-GroES complex. Interestingly, the SPM predicts the network of most probable residues that carry the signals. The network spans portions of the entire complex implying that all the regions participate during the mechanics of GroEL function. Interestingly, the residues with large stored elastic energy, which hold the key for allosteric signals, are also involved in the dynamics of allosteric transitions. Both from a structural and dynamic perspective there is an inherent asymmetry in the GroEL allostery with different subunits exhibiting distinct dynamics even though there is a seven fold symmetry in the taut state. 

We conclude with the following additional remarks.

\begin{itemize}
\item
There is an inherent asymmetry in the allosteric transitions. The SPM shows that  both the fluctuations and the stored elastic energies in residues belonging to adjacent E domains are drastically different.   This asymmetry is reflected in the dynamics of transition between distinct allosteric states. There are substantial molecule-to-molecule variations leading to heterogeneity in the allosteric transitions, which are hidden when ensemble averaging is performed. The implications, if any, of the heterogeneity  (found in molecular motors as well \cite{Hyeon07PNAS,Zhang2012Structure,Tehver10Structure}) for GroEL function is unclear. It should be emphasized, however, that the plasticity associated with the A domain \cite{Zheng07BJ,Fei13PNAS} could be relevant not only for SP recognition but also for ease of inter domain movement that is crucial in GroEL performing work on the SP \cite{Corsepius13PNAS}. 
Remarkably, the dispersal in the binding sites in the A domain during the $T \rightarrow R$ transition must occur as a result of the small torque exerted by the movements of key residues in the E domain through the I domain. The molecular basis of this form of allostery is well captured by the SPM and is also reflected in the dynamics. The estimated force experienced by the SP ($\sim (10 - 20)$ pN) is sufficient to partially unfold the SPs, especially considering that the domains move relatively slowly during the ATP-driven $T \rightarrow R^{\prime}$ transition.

\item
Our simulations show that the allosteric transitions are triggered by formation of salt-bridges, which are both of intra subunit and inter subunit variety. There is a clear hierarchy in the time scales in their rupture and formation \cite{Hyeon06PNAS}. More generally, it appears that in a variety of systems, including molecular switches, salt-bridges collectively drive allosteric transitions.  

\item 
 
We have not discussed the role of the dual cavity in GroEL, whose importance could be understood using the following arguments. In the IAM theory the larger $k_{R" \rightarrow T}$ is the more the yield of the native fold would be in biologically relevant time. 
A natural question arises: Has the GroEL machine evolved to maximize $k_{R" \rightarrow T}$? This is indeed the case. First, the residence time of a SP in the cavity is drastically smaller when a load, in the form of SP, is present. Second, ADP release from the {\it trans} ring in the symmetric complex is greatly accelerated in the presence of SP \cite{Ye13PNAS}, thus resetting the machine for yet another round of SP processing. This finding shows that when a load (SP) is imposed on GroEL it functions at an accelerated pace by going through the catalytic cycle rapidly (maximizing $k_{R"\rightarrow T}$), as anticipated by the IAM \cite{Todd96PNAS} and explicitly shown in \cite{Tehver08JMB}. The nearly hundred increase in SP-induced ADP release \cite{Ye13PNAS} is similar to the nearly thousand fold enhancement of ADP release from the motor head of kinesin in the presence of microtubule \cite{Hackney05PNAS}. Thus, just as molecular motors, GroEL functions as a genuine molecular machine sharing many common functional themes. 

\item
 A key recent development is that GroEL functions as a parallel processing machine in the sense that it can simultaneously process two SPs, one in each cavity \cite{Takei12JBC,Ye13PNAS}. From a structural perspective that the functional state, in the presence of SP, is the symmetric complex, which bears a close resemblance to the American football.   Thus, from this perspective also it follows that upon placing a load the GroEL machine turns over as rapidly as possible to maximize yield in biologically relevant time scale. In other words, it undergoes numerous cycles to achieve the functional objective, as described by the IAM. At the molecular level this requires signal transmission across several nanometers, which is achieved by multiple allosteric transitions.
 
The transition from $R^{\prime\prime} \rightarrow T$, which involves disassociation of GroES, is required to start a new cycle. If the symmetric complex is the functional state it poses a conundrum: Which of the two GroES particles bound to GroEL would   be dislodged first? It appears that the breakage of symmetry \cite{Ye13PNAS} in the inherently symmetric functional unit, crucial for maximizing the number of iterations in biologically relevant times, must be a dynamic process related to the differential in the number of ATP molecules hydrolyzed in the two rings. It will be most interesting to sort out the molecular basis of the communication that occurs over $\approx$ 16 nm!

\end{itemize}

{\bf Acknowledgements:} We are grateful to George Lorimer with whom many of the ideas, in particular the IAM, was developed. The contributions of Wenjun Zheng and Riina Tehver are gratefully acknowledged. This work was supported by a grant from the National Science Foundation (CHE 16-36424 and 16-32756) and the Collie-Welch Regents Chair (F-0019).

\clearpage

\clearpage 

\end{document}